\documentstyle[sprocl,epsfig]{article}

\def\be{\begin{equation}}
\def\ee{\end{equation}}
\def\bea{\begin{eqnarray}}
\def\eea{\end{eqnarray}}
\newcommand{\dst}{\displaystyle}

\newcommand{\epe}{\mbox{$e^+e^-\,$}}
\newcommand{\ggam}{\mbox{$\gamma\gamma\,$}}
\newcommand{\egam}{\mbox{$\gamma e\,$}}

\newcommand{\egeh}{\mbox{$e\gamma\to eH\,$}}

{\end{list}}
\newcounter{enumct}

\begin{document}

\begin{flushright}
IFT 99/18 \\
{\bf }
\end{flushright}
\vskip 0.5cm



\title{Distinguishing Higgs Models at Photon Colliders\footnote{Contribution 
to the Proceedings of the International Workshop 
on Linear Colliders, April 28-May 5, 1999, Sitges (Spain)}}

\author{Ilya F. Ginzburg}

\address{Sobolev Institute of Mathematics, Siberian Branch of RAS,
Prosp.\ ac.\ Koptyug, 4, 630090 Novosibirsk, Russia}

\author{Maria Krawczyk}

\address{Institute of Theoretical Physics, Warsaw University, Poland}

\author{Per Osland}
\address{Department of Physics, University of Bergen,
Allegt.\ 55, N-5007 Bergen, Norway}

\maketitle

\abstracts{
We consider the two effective couplings $h Z\gamma$ and $h \gamma \gamma$
involving a neutral scalar Higgs boson with a mass around 100 GeV
in the Standard Model, in the Two Higgs Doublet Model, and
in the Minimal Supersymmetric Standard Model.
The couplings can be tested at Photon Colliders, 
and used to distinguish these models.}
\section{Introduction}
Many recent studies for the TEVATRON, LHC and HERA assume, in fact, that
Nature is so favorable for us that new particles are sufficiently light
that they can be seen at these facilities.
We should like to discuss here the opposite scenario:
{\it No new particles will be discovered at these facilities,
except the Higgs scalar(s)}.

In this scenario, additional particles could very well exist
in addition to the Higgs boson, but they would have to be heavier 
than 1--2 TeV.
What new physics is realized 
could then be revealed at an $e^+e^-$ Linear Collider,\cite{Accomando}
where the {\it direct} couplings of the Higgs boson with matter will be
measured with high accuracy. Other couplings, which only occur
at the one-loop level, like $h\gamma\gamma$ and $hZ\gamma$, can only
be studied with higher statistics, and then only at lower accuracy.
The importance  of these couplings is related
to the fact that in the SM and in its extensions, all fundamental
charged particles are included in the loop, so  the structure of the 
theory influences  the corresponding Higgs boson decays.
The \ggam\ and \egam\ modes of a Linear Collider (Photon
Colliders) \cite{GKST} provide an opportunity to measure these vertices
with high enough accuracy, up to the 2\% level or better for the \ggam\
mode \cite{JikS} and up to a few percents for the $Z\gamma$ mode
(reaction \egeh). Therefore, the study of the Higgs boson production
at Photon Colliders could distinguish different models of new
physics prior to the discovery of other related new particles.

Frequently considered models going beyond the SM are 
the Two Higgs Doublet Model (2HDM), and
the Minimal Supersymmetric Standard Model (MSSM).
In the present paper we compare these loop-determined effective couplings 
$h\gamma\gamma$ and $hZ\gamma$, in the SM ($h=H_{\rm SM}$), 
in the 2HDM (Model II) where only the Higgs sector is enlarged 
compared to the SM, and in the MSSM, 
where the Higgs sector has formally the structure of Model II,
but where the parameters are more constrained, and 
where, in addition, new supersymmetric particles appear.  
We study properties of the couplings of the Higgs bosons with
photon(s) for the case when the mass of the lightest Higgs particle $h$
in the 2HDM and the MSSM is in the region 100--130 GeV, which is
still open for all three models.

These effective couplings are to a large
extent determined by the couplings of the Higgs particle to the $W$,
to the $b$ and $t$ quarks, and to the charged Higgs boson.\cite{Gunion}
In terms of the parameters $\beta$
(which parameterizes the ratio of the two vacuum expectation values)
and $\alpha$ (which parameterizes the mixing among the two neutral,
$CP$-even Higgs particles), these couplings are proportional to
\bea
g_{hWW}&\sim& \sin(\beta-\alpha) \nonumber \\
g_{hbb}&\sim& -\frac{\sin\alpha}{\cos\beta}
=\sin(\beta-\alpha)-\tan \beta \cos(\beta-\alpha) \nonumber \\
g_{htt}&\sim& \frac{\cos\alpha}{\sin\beta}
=\sin(\beta-\alpha)+\frac{\cos(\beta-\alpha)}{\tan \beta},
\label{eq:couplings}
\eea
as compared with the SM couplings. 

Within the scenario that no new physics, except the Higgs
particle(s), is discovered at hadron or \epe\ colliders, we can
imagine two cases considered in the subsequent sections.

\section{A light SM-like Higgs boson}

The first studied opportunity is as follows. 
{\it All
direct couplings are the same as in the SM with one Higgs
doublet. How do we then determine whether the SM, the 2HDM or
the MSSM is realized?}
The answer can be obtained from a precise study of
the two-photon Higgs width and the $h Z\gamma$ coupling at Photon
Colliders, $\gamma\gamma$ and $e\gamma$,\cite{GKST}
where the current estimate of the accuracy in the
measurement of the first width is of 2\%.\cite{JikS}
Indeed, in the SM, these vertices are determined by contributions from 
$W$ loops and matter loops, entering with opposite signs.\cite{Ellis}
Because of this partial cancellation, 
the addition of new contributions could change these vertices significantly.

This situation can be realized in the 2HDM, with $\beta -\alpha =\pi/2$, 
leading to couplings to gauge bosons and fermions as in the SM
(see above), and to some extent also in the MSSM.  
We have calculated the widths in these models for this case, as functions 
of $\tan \beta$, keeping for the 2HDM $\sin (\beta -\alpha) = 1$.
However, the MSSM with given $M_h$ and $\beta-\alpha$ values can be
realized only at some definite value of $\beta$ (if masses of
heavy squarks are roughly fixed), see Fig.~1.

For the 2HDM, the difference with respect to the SM is in this case
determined by the charged Higgs boson contribution only.
The relevant quantity becomes in this case 
$b=1+(M_h^2-\lambda_5v^2)/(2M^2_{H^\pm})$.\cite{Djouadi1} 
In the calculation for the general 2HDM(II) we assume 
$\lambda_5v^2\ll M_h^2$. 
This effect of the scalar loop is enhanced here due to
the partial compensation of the $W$ and $t$-quark
contributions. The result is evidently independent of the mixing
angle $\beta$. We find, for very heavy $H^\pm$ and $M_h$=100~GeV:
\be
\Gamma_{h\gamma\gamma}^{\rm 2HDM}/
\Gamma_{h\gamma\gamma}^{\rm SM} = 0.89, \qquad
\Gamma_{hZ\gamma}^{\rm 2HDM}/
\Gamma_{hZ\gamma}^{\rm SM} = 0.96.
\label{Eq:numbers}
\ee
The effect is of the order of about 10\%---a difference
large enough to be observed.
With growth of $\lambda_5$ this
effect decreases roughly as $(1-\lambda_5v^2/2M^2_{H^\pm})$.

In the MSSM we encounter two differences. First, with
a fixed mass of the lightest Higgs particle, only a finite range
of $\tan\beta$ is physical. Throughout this range, the coupling
of the lightest Higgs particle to the $W$ will vary 
as $\sin(\beta-\alpha)$, which is uniquely determined by $\tan\beta$
and the Higgs mass. Since this loop contribution dominates the
effective couplings under consideration, there will be a corresponding
strong variation with $\tan\beta$, as illustrated in Fig.~1.
Second, the contributions of the many superpartners depend on their masses. 
If all these masses are sufficiently heavy, the effects become 
small.\cite{Djou}

\section{Non-SM-like Higgs boson(s)}

The second possibility is that the couplings with matter differ
from those of the SM.  In this case it is very likely that this fact  is
known from earlier measurements, and our goal will be to search for an
opportunity to distinguish the cases of the 2HDM (II) and the MSSM.
In this respect we note that the measurements at a Linear Collider will give
us the couplings of the lightest Higgs boson with ordinary matter
and, perhaps, masses of some of the more heavy Higgs particles.
For a fixed mass $M_h$,
the $h\gamma\gamma$ and $h Z\gamma$ couplings are much smaller
in the 2HDM than in the MSSM, for a wide range of $\tan\beta$ values.
\section{Results}
We shall here present results for the
$h\to\gamma\gamma$ and $h\to Z\gamma$ widths
for a fixed Higgs boson mass equal to $M_h=100$~GeV.
The most recent values were taken for the fermion and gauge boson masses,
and other parameters.\cite{eps99}

In Fig.~1, we show the decay-rate ratios 
$\Gamma(h\to\gamma\gamma)/\Gamma(H^{\rm SM}\to\gamma\gamma)$ and
$\Gamma(h\to Z\gamma)/\Gamma(H^{\rm SM}\to Z\gamma)$, 
for the 2HDM and the MSSM.
For the 2HDM, we take $\sin(\beta-\alpha)=0$ and $1$, and
consider a range of  values of the charged Higgs boson mass
from 165~GeV to infinity.
Let us first discuss the case of $\sin(\beta-\alpha)=0$.
It is important to note that in this case, the lightest Higgs particle
decouples from the $W$ loops (see Eq.~(\ref{eq:couplings})).
Thus, the decay rates are dominantly due to $b$ and $t$ quark loops
(plus a small contribution from the charged Higgs particle).
Dips appear for $\tan\beta\simeq$~4--5, 
where the $b$-quark starts to dominate over the $t$-quark contribution.
The horizontal lines correspond to the
2HDM results for the case $\sin(\beta-\alpha)=1$ 
(see the numbers above).

\setcounter{figure}{0}
\begin{figure}
\begin{center}
\setlength{\unitlength}{1cm}
\begin{picture}(7.0,6.0)
\put(-2.7,0.0){\mbox{\epsfysize=6.3cm\epsffile{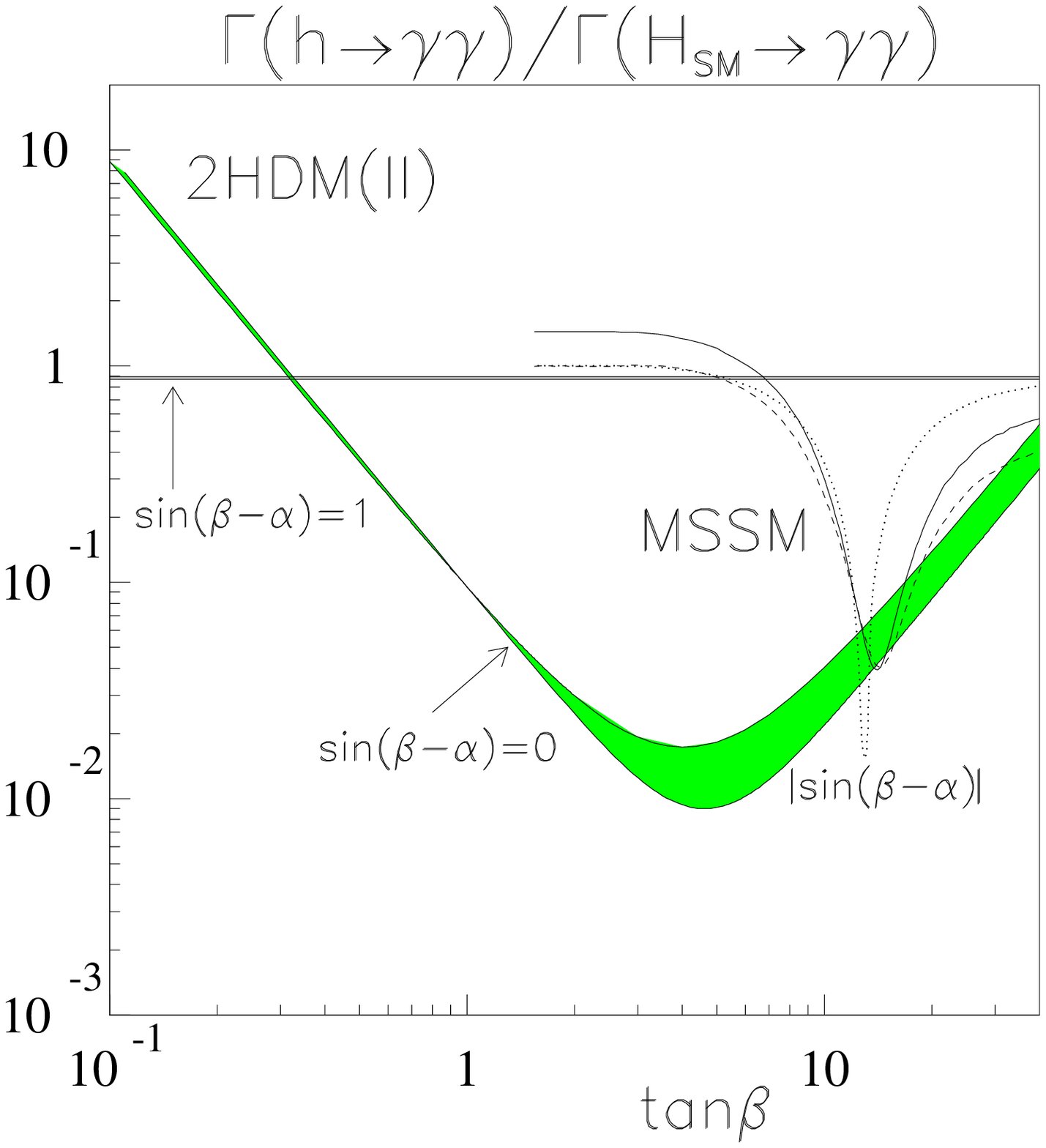}}
               \mbox{\epsfysize=6.3cm\epsffile{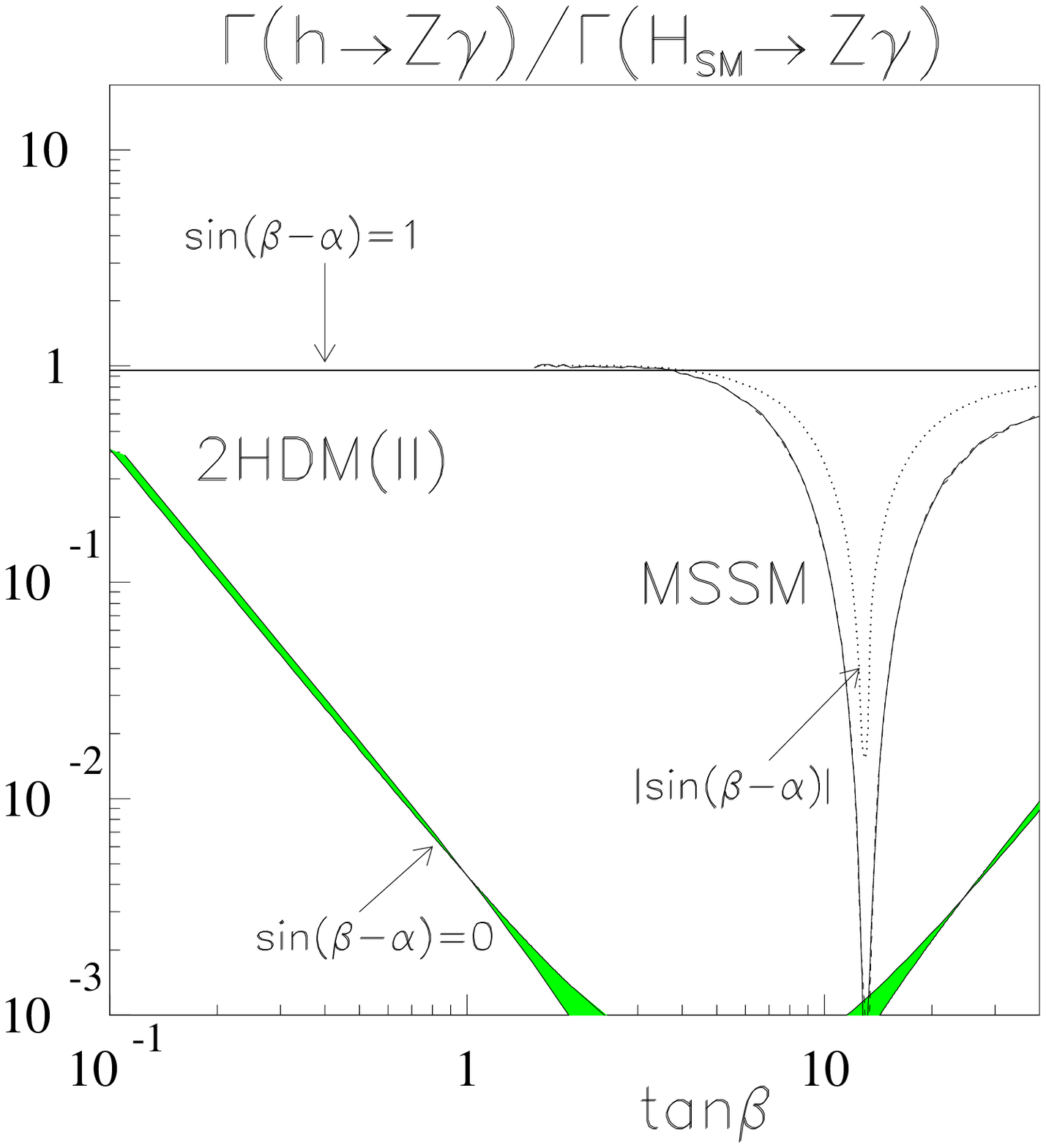}}}
\end{picture}
\caption{Ratios of the Higgs boson $h\to\gamma\gamma$ 
and $h\to Z\gamma$ decay widths in the 2HDM and the SM 
and between the MSSM and the SM as functions of $\tan \beta$.
Results correspond to the Higgs mass $M_h$=100 GeV.
In the 2HDM, the band for $\sin(\beta-\alpha)=0$ corresponds
to the mass of the charged Higgs boson ranging from 165 GeV to infinity.
For the MSSM curves, the solid ones include effects of 
supersymmetric loop particles (default masses), 
whereas the dashed ones do not. The dotted curve describes the function 
$|\sin(\beta-\alpha)|$ as a function of $\tan \beta$
for fixed mass $M_h$=100 GeV in the MSSM.}
\end{center}
\end{figure}

For the MSSM, 
we have used the results of the program {\tt HDECAY}.\cite{hdecay}
The solid (dashed) curves correspond to supersymmetric particles 
contributing (or not) to the loops.
Interestingly, 
for the $h\to Z\gamma$ decay, these two options are indistinguishable.
The sharp dips in these ratios at $\tan\beta\sim$ 12--14 correspond
to the vanishing of $|\sin(\beta-\alpha)|$
(shown separately as a dotted curve), which determines the $hWW$ coupling. 
 In contrast to the 2HDM, this coupling is {\it not} a free parameter 
in the MSSM, since the Higgs mass here is kept fixed.

As mentioned above, for the MSSM, low values of $\tan\beta$ 
are not physical. Where the curves terminate at low $\tan\beta$,
the charged Higgs mass is of the order of $10^5$~GeV.
The results discussed above were obtained for $M_h=100$~GeV.
We have checked that a similar picture holds for $M_h=120$~GeV.

In summary, we have shown that the Higgs couplings involving one
or two photons, which can be explored in detail at Photon Colliders
($\gamma\gamma$ and $e\gamma$),
could resolve the models SM, 2HDM or MSSM with similar neutral Higgs 
boson masses in the range $M_h\sim$~100--120~GeV and having similar
couplings to matter.
In the case of different coupling to matter, a clear distinction 
can be made between the 2HDM and the MSSM.
\medskip

This research has been supported by 
RFBR grants 99-02-17211 and 96-15-96030, by
Polish Committee for Scientific Research, grant No.\ 2P03B01414, 
and by the Research Council of Norway.

\end{document}